\documentclass[conference]{IEEEtran}
\IEEEoverridecommandlockouts

\usepackage{cite}
\usepackage{amsmath,amssymb,amsfonts}
\usepackage{graphicx}
\usepackage{textcomp}
\usepackage{xcolor}
\usepackage{algorithm}
\usepackage{algpseudocode}
\usepackage{upgreek}
\usepackage{multirow}
\def\BibTeX{{\rm B\kern-.05em{\sc i\kern-.025em b}\kern-.08em
    T\kern-.1667em\lower.7ex\hbox{E}\kern-.125emX}}

\begin{document}

\title{Distributed Droop-Free Control of Grid-Forming Inverters with Dynamic Line and Load Modeling*\\
\thanks{This work was supported by the MIT--GE Vernova Energy and Climate Alliance under the project “Energy Dynamics Modeling and Control Approach for Grid-Forming Electricity Services” (PI: M. Ilic). Hiya Gada is supported by the Department of Aeronautics and Astronautics Distinguished Scholars Program at the Massachusetts Institute of Technology. 
Accepted for publication at IEEE SmartGridComm 2026.}
}

\author{\IEEEauthorblockN{Hiya Akhil Gada}
\IEEEauthorblockA{\textit{Department of Aeronautics and Astronautics} \\
\textit{Massachusetts Institute of Technology}\\
Cambridge, USA \\
hiyagada@mit.edu}
\and
\IEEEauthorblockN{Marija Ilic}
\IEEEauthorblockA{\textit{Department of Electrical Engineering and Computer Science} \\
\textit{Massachusetts Institute of Technology}\\
Cambridge, USA \\
ilic@mit.edu}
}

\maketitle

\begin{abstract}
Droop-free distributed control has emerged as a promising alternative to conventional linear droop control for coordinating inverter-based resources in AC microgrids.
However, existing droop-free methods typically rely on quasi-steady state network models that neglect fast electromagnetic transients and assume a decoupled dependence of active and reactive power on frequency and voltage, respectively.
In this paper, we propose a transient-aware droop-free distributed primary/secondary control framework for grid-forming (GFM) inverters.
The control objective is to achieve proportional active and reactive power sharing, frequency regulation, and voltage regulation within prescribed bounds.
The proposed approach incorporates dynamic models of network lines and loads into the control design, enabling accurate representation of transient behavior while capturing the coupled nature of power flow interactions.
A gradient-based distributed control update is derived from an optimization formulation, in which voltage magnitude constraints are enforced through a projection operator.
To validate the performance of the proposed control, real-time simulation studies are conducted on a four-inverter microgrid under load changes in both strong and weak grid conditions.
The results demonstrate that the proposed method improves transient performance, particularly in weak grids, while preserving the same steady state operating conditions as existing QSS-based droop-free approaches.
\end{abstract}


\section{Introduction}\label{sec:intro}

Microgrids are small-scale power systems that integrate distributed energy resources, enabling clusters of generation, storage, and loads to operate either grid-connected or in islanded mode~\cite{lasseter2002}.
In such systems, power electronic inverters act as the interface between energy sources and the AC network, and are typically operated in either grid-following (GFL) or grid-forming (GFM) modes.
GFL inverters synchronize to an externally imposed voltage and frequency and regulate their power injection accordingly, whereas GFM inverters establish the voltage waveform and frequency~\cite{bahrani2024, miller2024}.

A hierarchical control architecture is commonly adopted for microgrid operation, consisting of primary, secondary, and tertiary control layers~\cite{guerrero2011,bidram2012}. 
Primary control is responsible for voltage and frequency stability while ensuring proportional power sharing among sources, secondary control compensates for the steady state deviations introduced by primary control, and
tertiary control performs economic dispatch over the network. 
Primary control is traditionally implemented using droop control~\cite{chandorkar}, a decentralized strategy that emulates the steady state behavior of synchronous generators to achieve proportional active and reactive power sharing~\cite{pogaku2007}.

Despite its simplicity, scalability, and its ability to operate without communication between sources, droop control has fundamental limitations. 
First, the use of linear droop does not guarantee accurate power sharing under nonlinear loading conditions~\cite{nonlinearload} and introduces steady state errors in frequency and voltage as loads vary~\cite{haddadi2014}. 
Second, conventional droop assumes predominantly inductive networks, allowing decoupling of active power–frequency and reactive power–voltage dynamics~\cite{bidram2012,inductiveload}.
Finally, accurate reactive power sharing with droop is sensitive to line impedance mismatches and unequal bus voltages, since voltage is a local variable and not synchronized across the network, unlike frequency~\cite{shafiee2014}.

Thus, several improved droop-based control, such as adaptive droop~\cite{adaptive1,adaptive2}, virtual flux droop~\cite{fluxdroop} and virtual impedance methods~\cite{virtualimp1,virtualimp2}, have been proposed to mitigate some of these limitations.
However, these approaches do not fully resolve the aforementioned issues, particularly those arising from coupling between active and reactive power dynamics. 
%
This has motivated the development of droop-free distributed control approaches~\cite{nasirian2016,mohiuddinunified,mohiuddin2020}, where inverters coordinate with neighboring units over a sparse communication network to achieve improved regulation and power sharing without relying on droop characteristics.
These methods can approximate centralized performance while maintaining scalability, robustness, and low computational complexity.

Nevertheless, existing droop-free methods still rely on simplifying assumptions that limit their applicability.
The foundational work in~\cite{nasirian2016} adopts a quasi-steady state (QSS) network impedance model, where line and load dynamics are approximated by algebraic voltage–current relations, neglecting fast electromagnetic transients relevant in inverter-dominated systems. 
Moreover, it assumes a decoupled dependence of active and reactive power on frequency and voltage, respectively.
In~\cite{mohiuddinunified}, a unified droop-free architecture is proposed for both GFL and GFM inverters, achieving frequency regulation and average voltage tracking through a distributed optimization formulation. 
However, regulating only the global average voltage does not ensure individual bus voltages remain within admissible limits~\cite{mohiuddin2020}. 
To address this,~\cite{mohiuddin2020} introduces a voltage variance regulator to enforce voltage profile constraints, but at the cost of excluding one inverter, typically the largest-capacity unit, from reactive power sharing, leading to operational imbalance.
More fundamentally, both approaches continue to rely on QSS network models and do not fully capture transient dynamics or coupled power interactions.

To address these limitations, this paper proposes a transient-aware droop-free distributed primary/secondary control framework for GFM inverters. 
The main contributions are:

\begin{enumerate}

\item A dynamic network representation that models line and load currents using differential equations, rather than QSS algebraic approximations, capturing fast electromagnetic transients and the coupled dependence of active and reactive power on voltage and frequency without assuming inductive networks.

\item A droop-free gradient-based distributed control framework derived from a distributed optimization problem, where gradients explicitly incorporate current sensitivities to local voltage from the dynamic network model, improving transient performance under load changes (particularly in weak grids) while requiring only local measurements and limited neighbor communication. 

\item A projection operator that enforces voltage magnitude constraints while preserving participation of all units in power sharing, ensuring normalized powers remain as closely matched as possible without excluding inverters from the sharing objective or relying solely on average voltage regulation.

\end{enumerate}

The remainder of the paper is organized as follows. 
Section~\ref{sec:gfm_control} provides an overview of the zero-level control architecture of GFM inverters.
Section~\ref{sec:droop_free} presents the proposed transient-aware droop-free distributed control framework, including the communication model, problem formulation, distributed gradient computation, and projected gradient-based control design, along with steady state performance analysis. 
Section~\ref{sec:results} provides simulation results demonstrating the effectiveness of the proposed approach under both strong and weak grid conditions, highlighting improvements in transient performance relative to QSS-based methods. 
Finally, Section~\ref{sec:conclusion} concludes the paper and outlines directions for future work.

\section{Zero-level control of GFM inverters}\label{sec:gfm_control}
\begin{figure}
    \centering
    \includegraphics[width=\linewidth]{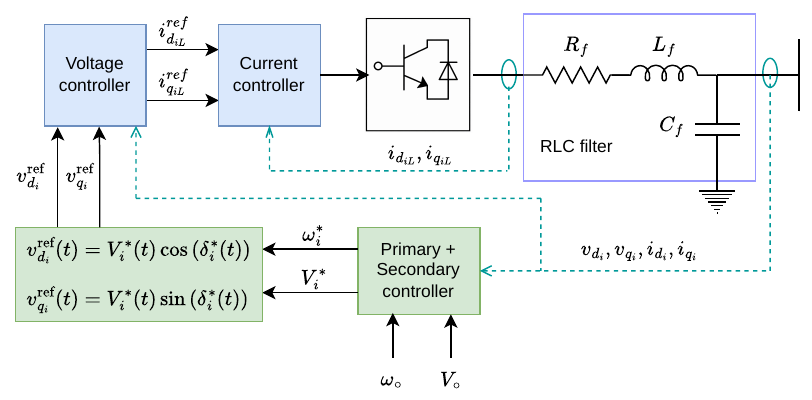}
    \caption{GFM inverter with cascaded voltage and current control loops and RLC filter.}
    \label{fig:gfm_control}
\end{figure}
A GFM inverter operates as a controllable voltage source that generates a voltage waveform at its terminal by tracking the frequency and voltage magnitude setpoints provided by primary/secondary control. 

As shown in Fig.~\ref{fig:gfm_control}, the zero-level control of a GFM inverter with an $RLC$ filter consists of cascaded inner voltage and current control loops implemented in a $dq$ reference frame rotating at the nominal angular frequency $\omega_\circ$. 
Primary control operates on a faster timescale than secondary control (traditionally via droop) to regulate power sharing, while secondary control corrects steady state deviations in voltage and frequency.
Together, they provide the voltage magnitude and frequency references, $V_i^*(t)$ and $\omega_i^*(t)$, which define the $dq$ voltage references as,
\begin{align}
    v_{d_i}^{\text{ref}}(t) &= V_i^*(t)\cos\left(\delta_i^*(t)\right), \label{eqn:vd}\\
    v_{q_i}^{\text{ref}}(t) &= V_i^*(t)\sin\left(\delta_i^*(t)\right), \label{eqn:vq}
\end{align}
where the phase angle $\delta_i^*$ evolves according to,
\begin{align}\label{eqn:phase_angle}
    \delta_i^*(t) = \int_0^t \left(\omega_i^*(\tau) - \omega_\circ\right) d\tau.
\end{align}
The cascaded voltage and current control loops track these references, enabling the inverter to behave as a controlled voltage source at the grid terminal.

\section{Transient-aware droop-free distributed control of GFM inverters}\label{sec:droop_free}

\subsection{Distributed communication model}
For simplicity, we consider a microgrid network consisting only of inverter output buses. 
We model the electrical network as an undirected graph $\mathcal{G} = (\mathcal{B}, \mathcal{E})$, where $\mathcal{B}$ denotes the set of GFM inverters and $\mathcal{E}$ denotes the set of electrical interconnections. 
For each inverter $i \in \mathcal{B}$, let $\mathcal{M}_i:=\{j \in \mathcal{B} \mid (i,j) \in \mathcal{E} \}$ denote the set of its \emph{electrical neighbors}, i.e., inverters directly connected to $i$ through network lines.

In addition, we define a directed communication graph $\mathcal{N} = (\mathcal{B}, \mathcal{E}_c)$, where $\mathcal{E}_c$ denotes the set of communication links. 
For each inverter $i \in \mathcal{B}$, let $\mathcal{N}_i := \{ j \in \mathcal{B} \mid (j,i) \in \mathcal{E}_c \}$ denote the set of its \emph{in-neighbors} (predecessors), i.e., inverters from which $i$ receives information.
The communication graph $\mathcal{N}$ is assumed to contain a spanning tree, ensuring information propagation across the entire network.

Under this architecture, each inverter $i$ receives from its communication neighbors $j \in \mathcal{N}_i$ their respective active and reactive power injections $(P_j, Q_j)$ and $dq$-frame voltage measurements $(v_{d_j}, v_{q_j})$. 
All other quantities required for control implementation, including local voltages, currents, and line and load parameters, are assumed to be available locally.

For notational simplicity, the superscript ``ref'' is omitted in the sequel, and $v_{d_i}, v_{q_i}$ denote the voltage reference signals generated by the proposed droop-free distributed control.

\subsection{Problem formulation}\label{subsec:problem_formulation}
The objective of the proposed droop-free distributed control is to coordinate multiple GFM inverters in an AC microgrid such that (i) active and reactive power are shared in proportion to inverter capacities, (ii) the network frequency is regulated to the nominal value $\omega_\circ$, and (iii)~inverter terminal voltages are regulated within prescribed bounds.

We formulate the following optimization problem:
\begin{align*}
    \min_{\{v_{d_i},\, v_{q_i}\}_{i \in \mathcal{B}}} \quad 
    & J = \sum_{i \in \mathcal{B}} J_i \\
    \text{s.t.} \quad 
    & \underline{V}^2 \leq v_{d_i}^2 + v_{q_i}^2 \leq \overline{V}^2, \quad \forall i \in \mathcal{B},
\end{align*}
where the local objective function for inverter $i \in \mathcal{B}$ is defined as
\begin{align*}
    J_i = 
    \frac{\beta_p}{2} \sum_{j \in \mathcal{N}_i} \left(P_i^n - P_j^n\right)^2 
    + \frac{\beta_q}{2} \sum_{j \in \mathcal{N}_i} \left(Q_i^n - Q_j^n\right)^2.
\end{align*}
Here, $\beta_p, \beta_q > 0$ are design parameters, and $P_i^n$, $Q_i^n$ denote the normalized active and reactive power injections, 
\begin{align*}
    P_i^n = \frac{P_i}{\overline{P}_i}, \quad 
    Q_i^n = \frac{Q_i}{\overline{Q}_i},
\end{align*}
where $\overline{P}_i$ and $\overline{Q}_i$ are the rated active and reactive power capacities of inverter $i$.

\subsection{Distributed gradient computation} 
For each inverter $i \in \mathcal{B}$, the active and reactive power injections in the $dq$ reference frame are given by
\begin{align}
P_i &= v_{d_i} i_{d_i} + v_{q_i} i_{q_i}, \label{eqn:power} \\
Q_i &= v_{q_i} i_{d_i} - v_{d_i} i_{q_i}. \label{eqn:reac_power}
\end{align}

The gradient of $J_i$ with respect to $\mathbf{v}_i$ is given by
\begin{align}\label{eqn:main_gradient}
\nabla_{\mathbf{v}_i} J_i
&=
\sum_{j \in \mathcal{N}_i}
\Big[
\beta_p (P_i^n - P_j^n)\left(\frac{\nabla_{\mathbf{v}_i} P_i}{\overline{P}_i} - \frac{\nabla_{\mathbf{v}_i} P_j}{\overline{P}_j}\right) \nonumber \\
& \hspace{1cm}
+
\beta_q (Q_i^n - Q_j^n)\left(\frac{\nabla_{\mathbf{v}_i} Q_i}{\overline{Q}_i} - \frac{\nabla_{\mathbf{v}_i} Q_j}{\overline{Q}_j}\right)
\Big],
\end{align}
where $\mathbf{v}_i := [v_{d_i} \;\; v_{q_i}]^\top$.

To evaluate $\nabla_{\mathbf{v}_i} P_i$, $\nabla_{\mathbf{v}_i} Q_i$, $\nabla_{\mathbf{v}_i} P_j$, and $\nabla_{\mathbf{v}_i} Q_j$, the partial derivatives of the inverter currents with respect to the local voltage $\mathbf{v}_i$ are required. 
To characterize this dependence, we express the inverter output current as the sum of line and load currents. Specifically,
\begin{align*}
\mathbf{i}_i = \sum_{j \in \mathcal{M}_i} \mathbf{i}_{ij} + \mathbf{i}_{\ell i},
\end{align*}
where $\mathbf{i}_{ij}$ denotes the current flowing from inverter $i$ to its electrical neighbor $j$, and $\mathbf{i}_{\ell i}$ denotes the local load current at bus $i$, both expressed in the $dq$ frame.
In many existing droop-free approaches~\cite{nasirian2016,mohiuddinunified,mohiuddin2020}, this dependence is approximated using a quasi-steady state (QSS) network impedance model, yielding algebraic voltage-current relations as,
\begin{align}
\frac{\partial \mathbf{i}_{ij}}{\partial \mathbf{v}_i}
&= \frac{1}{R_{ij}^2 + (\omega_\circ L_{ij})^2}
\left( R_{ij}\mathbf{I}_2 - \omega_\circ L_{ij}\mathbf{J} \right), \label{eqn:Iij_QSS} \\
\frac{\partial \mathbf{i}_{\ell i}}{\partial \mathbf{v}_i}
&= \frac{1}{R_{\ell i}^2 + (\omega_\circ L_{\ell i})^2}
\left( R_{\ell i}\mathbf{I}_2 - \omega_\circ L_{\ell i}\mathbf{J} \right). \label{eqn:Ili_QSS}
\end{align}
However, such approximations neglect fast electromagnetic transients, which are significant in inverter-dominated microgrids.

In this work, we propose a transient-aware controller, wherein the gradient computation explicitly accounts for the dynamics of both network lines and local loads through their ODE representations, as opposed to the algebraic relations in~\eqref{eqn:Iij_QSS} and \eqref{eqn:Ili_QSS}.
Consider an $RL$ line connecting inverter $i$ to a neighbor $j$. 
The line current dynamics in the $dq$ frame are given by
\begin{align}\label{eqn:RL_ij}
L_{ij} \frac{{d\mathbf{i}}_{ij}}{dt}
=
- R_{ij} \mathbf{i}_{ij}
+ \mathbf{v}_i - \mathbf{v}_j
- \omega_\circ L_{ij} \mathbf{J} \mathbf{i}_{ij},
\end{align}
where
\[
\mathbf{J} =
\begin{bmatrix}
0 & -1 \\
1 & 0
\end{bmatrix}.
\]

For a local time-varying $RL$ load connected at inverter $i$, the load current satisfies
\begin{align}\label{eqn:RL_load}
L_{\ell i} \frac{{d\mathbf{i}}_{\ell i}}{dt} 
=
- (R_{\ell i} + \dot{L}_{\ell i}) \mathbf{i}_{\ell i}
+ \mathbf{v}_i
- \omega_\circ L_{\ell i} \mathbf{J} \mathbf{i}_{\ell i}.
\end{align}

Taking the partial differential of~\eqref{eqn:RL_ij} with $\mathbf{v}_i$ yields 
\begin{align} \label{eqn:Iij} 
\frac{d}{dt}\left(\frac{\partial \mathbf{i}_{ij}}{\partial \mathbf{v}_i}\right) = - \frac{R_{ij}}{L_{ij}} \frac{\partial \mathbf{i}_{ij}}{\partial \mathbf{v}_i} + \frac{1}{L_{ij}}\mathbf{I}_2 - \omega_\circ \mathbf{J} \frac{\partial \mathbf{i}_{ij}}{\partial \mathbf{v}_i}. 
\end{align}

Similarly, taking the partial differential of~\eqref{eqn:RL_load} gives
\begin{align} \label{eqn:Ili}
\frac{d}{dt}\left(\frac{\partial \mathbf{i}_{\ell i}}{\partial \mathbf{v}_i}\right) = - \frac{R_{\ell i} + \dot{L}_{\ell i}}{L_{\ell i}} \frac{\partial \mathbf{i}_{\ell i}}{\partial \mathbf{v}_i} + \frac{1}{L_{\ell i}}\mathbf{I}_2 - \omega_\circ \mathbf{J} \frac{\partial \mathbf{i}_{\ell i}}{\partial \mathbf{v}_i}. 
\end{align}

The dynamics in~\eqref{eqn:Iij} and~\eqref{eqn:Ili} define linear time-varying systems, which can be solved recursively in time, initialized with known conditions $\frac{\partial \mathbf{i}_{ij}}{\partial \mathbf{v}_i}(0) = \mathbf{h}_{ij}^\circ$ and $\frac{\partial \mathbf{i}_{\ell i}}{\partial \mathbf{v}_i}(0) = \mathbf{h}_{\ell i}^\circ$. 
In the case of constant line and load parameters, the steady state solutions of~\eqref{eqn:Iij} and~\eqref{eqn:Ili} reduce to the algebraic expressions in~\eqref{eqn:Iij_QSS} and~\eqref{eqn:Ili_QSS}.

The total current sensitivity to the local voltage at inverter $i$ is then given by,
\begin{align}\label{eqn:current_sens}
\frac{\partial \mathbf{i}_{i}}{\partial \mathbf{v}_i}
=
\sum_{j \in \mathcal{M}_i}
\frac{\partial \mathbf{i}_{ij}}{\partial \mathbf{v}_i}
+
\frac{\partial \mathbf{i}_{\ell i}}{\partial \mathbf{v}_i}.
\end{align}

Using the power expressions~\eqref{eqn:power}, \eqref{eqn:reac_power} and the current sensitivities obtained from~\eqref{eqn:Iij}–\eqref{eqn:current_sens}, the gradients of $P_i$, $Q_i$, $P_j$, and $Q_j$ can be expressed as
\begin{align}
\nabla_{\mathbf{v}_i} P_i
&= \begin{bmatrix}
i_{d_i} \\
i_{q_i}
\end{bmatrix}
 +
\begin{bmatrix}
v_{d_i} & v_{q_i}
\end{bmatrix}
\frac{\partial \mathbf{i}_{i}}{\partial \mathbf{v}_i}, \label{eqn:grad_P}\\
\nabla_{\mathbf{v}_i} Q_i
&=
\begin{bmatrix}
- i_{q_i} \\
i_{d_i}
\end{bmatrix}
+
\begin{bmatrix}
v_{q_i} & -v_{d_i}
\end{bmatrix}
\frac{\partial \mathbf{i}_{i}}{\partial \mathbf{v}_i}, \label{eqn:grad_Q} \\
\nabla_{\mathbf{v}_i} P_j
&= -\begin{bmatrix}
v_{d_j} & v_{q_j}
\end{bmatrix}
\frac{\partial \mathbf{i}_{ij}}{\partial \mathbf{v}_i}, \label{eqn:grad_Pj}\\
\nabla_{\mathbf{v}_i} Q_j 
&= -\begin{bmatrix}
v_{q_j} & -v_{d_j}
\end{bmatrix} 
\frac{\partial \mathbf{i}_{ij}}{\partial \mathbf{v}_i}. \label{eqn:grad_Qj}
\end{align}

\paragraph*{Remark}
It can be observed that the above gradients depend only on locally available quantities, including $(\mathbf{v}_i, \mathbf{i}_i)$, line and load $RL$ parameters, and communicated neighboring voltage and power measurements $(\mathbf{v}_j, P_j, Q_j)$ for $j \in \mathcal{N}_i$, preserving the distributed structure of the control design.

\subsection{Projected gradient descent}

The proposed transient-aware droop-free controller is obtained by substituting~\eqref{eqn:grad_P}--\eqref{eqn:grad_Qj} into~\eqref{eqn:main_gradient} and incorporating a projected gradient descent update for the voltage reference at each inverter.
\begin{align}\label{eqn:pgd_update}
\mathbf{v}_i(t + \Delta t)
=
\Pi_{\mathcal{V}}
\left[
\mathbf{v}_i(t)
-
\alpha_i \, \nabla_{\mathbf{v}_i} J_i(t)
\right],
\end{align}
where $\alpha_i > 0$ is the local step size, and $\Pi_{\mathcal{V}}(\cdot)$ denotes the projection onto the feasible voltage set
\begin{align*}
\mathcal{V}
=
\left\{
\mathbf{v}_i \in \mathbb{R}^2 \; \middle| \;
\underline{V}^2 \leq \|\mathbf{v}_i\|^2 \leq \overline{V}^2
\right\}.
\end{align*}

The projection operator ensures that the updated voltage reference satisfies the prescribed magnitude constraints. In particular, the projection onto $\mathcal{V}$ admits a closed-form expression given by
\begin{align}
\Pi_{\mathcal{V}}(\mathbf{v}_i)
=
\begin{cases}
\displaystyle
\overline{V} \, \frac{\mathbf{v}_i}{\|\mathbf{v}_i\|}, & \|\mathbf{v}_i\| > \overline{V}, \\[8pt]
\displaystyle
\underline{V} \, \frac{\mathbf{v}_i}{\|\mathbf{v}_i\|}, & \|\mathbf{v}_i\| < \underline{V}, \\[8pt]
\mathbf{v}_i, & \text{otherwise}.
\end{cases}
\end{align}

\paragraph*{Continuous-time implementation}
The discrete-time update in~\eqref{eqn:pgd_update} admits the following continuous-time limit:
\begin{align}
\dot{\mathbf{v}}_i
=
- \alpha_i \, \nabla_{\mathbf{v}_i} J_i,
\end{align}
with projection onto the feasible set $\mathcal{V}$.

\subsection{Steady state performance analysis}
 
At steady state, the voltage reference dynamics satisfy
\begin{align*}
\dot{\mathbf{v}}_i^{ss} = \mathbf{0}, \quad \forall i \in \mathcal{B}.
\end{align*}
If the voltage magnitude constraints are inactive at steady state, then from~\eqref{eqn:pgd_update}, this implies
\begin{align*}
\nabla_{\mathbf{v}_i} J_i^{ss} = \mathbf{0}, \quad \forall i \in \mathcal{B}.
\end{align*}

Under this condition, normalized active and reactive power injections are equal across neighboring inverters, i.e.,
\begin{align*}
P_i^{n,ss} = P_j^{n,ss}, \quad Q_i^{n,ss} = Q_j^{n,ss}, \quad \forall j \in \mathcal{N}_i.
\end{align*}
If the communication graph $\mathcal{N}$ contains a spanning tree, this pairwise agreement extends across the entire network, thereby enforcing proportional power sharing among all inverters.

Next, we examine frequency regulation. Using~\eqref{eqn:vd}--\eqref{eqn:phase_angle}, the frequency deviation at steady state from nominal is given by
\begin{align*}
\omega_i^{ss} - \omega_\circ 
&= \frac{d}{dt} \arctan\left(\frac{v_{q_i}^{ss}}{v_{d_i}^{ss}}\right), \\
&= \frac{v_{d_i}^{ss}\dot{v}_{q_i}^{ss} - v_{q_i}^{ss}\dot{v}_{d_i}^{ss}}{(v_{d_i}^{ss})^2 + (v_{q_i}^{ss})^2}.
\end{align*}
Since $\dot{v}_{d_i}^{ss} = 0$ and $\dot{v}_{q_i}^{ss} = 0$, it follows that
\begin{align*}
\omega_i^{ss} = \omega_\circ, \quad \forall i \in \mathcal{B}.
\end{align*}

Finally, the projection operator ensures that the voltage magnitude constraints are satisfied at all times, and hence at steady state,
\begin{align*}
\underline{V} \leq \|\mathbf{v}_i^{ss}\| \leq \overline{V}, \quad \forall i \in \mathcal{B}.
\end{align*}

In summary, the proposed control achieves at steady state: (i) proportional active and reactive power sharing when voltage constraints are inactive (and otherwise the closest feasible sharing under the constraints), (ii) frequency regulation to $\omega_\circ$, and (iii) voltage regulation within prescribed bounds.

\section{Results}\label{sec:results}

\begin{figure}
    \centering
    \includegraphics[width=0.9\linewidth]{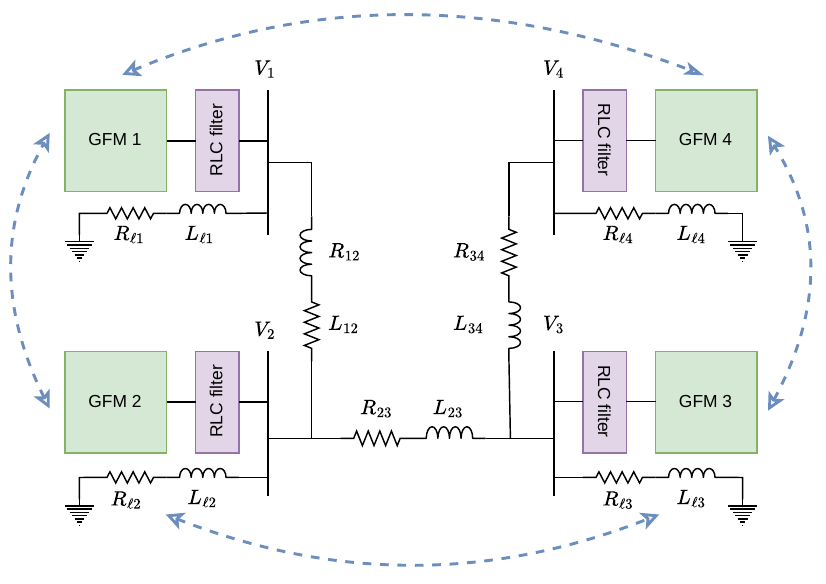}
    \caption{Radial microgrid with four GFM inverters. Blue dashed lines denote communication links.}
    \label{fig:MG}
\end{figure}

Two experiments are performed to evaluate the proposed transient-aware droop-free distributed control.
A schematic of the experimental setup is shown in Fig.~\ref{fig:MG}, comprising a radial microgrid with four GFM inverters and an associated communication graph represented by blue dashed lines.
Each inverter supplies a local load, and the system is initialized at a feasible steady state satisfying both equal power sharing and voltage constraints. 

The rated voltage of the microgrid is taken as $V_\circ = 208 \frac{\sqrt{2}}{\sqrt{3}}$.
Each GFM inverter is interfaced through an RLC filter with parameters $R_f = 0.1 \, \Omega$, $L_f =1.8 \, \text{mH}$, and $C_f = 25 \, \upmu \text{F}$.
The active and reactive power capacities are chosen as $\overline{P}_i = 600 \, \text{W}$ and $\overline{Q}_i = 600 \, \text{VAr}$ for $i = 1,2,3$, and $\overline{P}_4 = 1200 \, \text{W}$ and $\overline{Q}_4 = 1200 \, \text{VAr}$.
The objective function weights are selected as $\beta_p = 10$ and $\beta_q = 10$.

\begin{table}
\centering
\caption{Line Parameters}
\label{tab:line_params}
\begin{tabular}{|c|c|c|c|c|}
\hline
 & Line & 1--2 & 2--3 & 3--4 \\
\hline \hline
\multirow{2}{*}{Study 1} & $R~(\Omega)$ & 0.8 & 0.4 & 0.6 \\
\cline{2-5}
                       & $L~(\mathrm{mH})$ & 10 & 5 & 8 \\
\hline
\multirow{2}{*}{Study 2} & $R~(\Omega)$ & 0.4 & 0.4 & 0.4 \\
\cline{2-5}
                       & $L~(\mathrm{mH})$ & 50 & 50 & 50 \\
\hline
\end{tabular}
\end{table}

\begin{table}
\centering
\caption{Load Parameters}
\label{tab:load_params}
\begin{tabular}{|c|c|c|c|c|c|}
\hline
 & Load & $\ell 1$ & $\ell 2$ & $\ell 3$ & $\ell 4$ \\
\hline \hline
\multirow{2}{*}{Study 1} & $R~(\Omega)$ & 50 & 50 & 80 & 80 \\
\cline{2-6}
                       & $L~(\mathrm{mH})$ & 100 & 100 & 200 & 200 \\
\hline
\multirow{2}{*}{Study 2} & $R~(\Omega)$ & 0.8 & 0.8 & 0.8 & 0.8 \\
\cline{2-6}
                       & $L~(\mathrm{mH})$ & 100 & 20 & 20 & 20 \\
\hline
\end{tabular}
\end{table}

\subsection{Study 1: Performance in a strong grid}

\begin{figure}
    \centering
    \includegraphics[width=\linewidth]{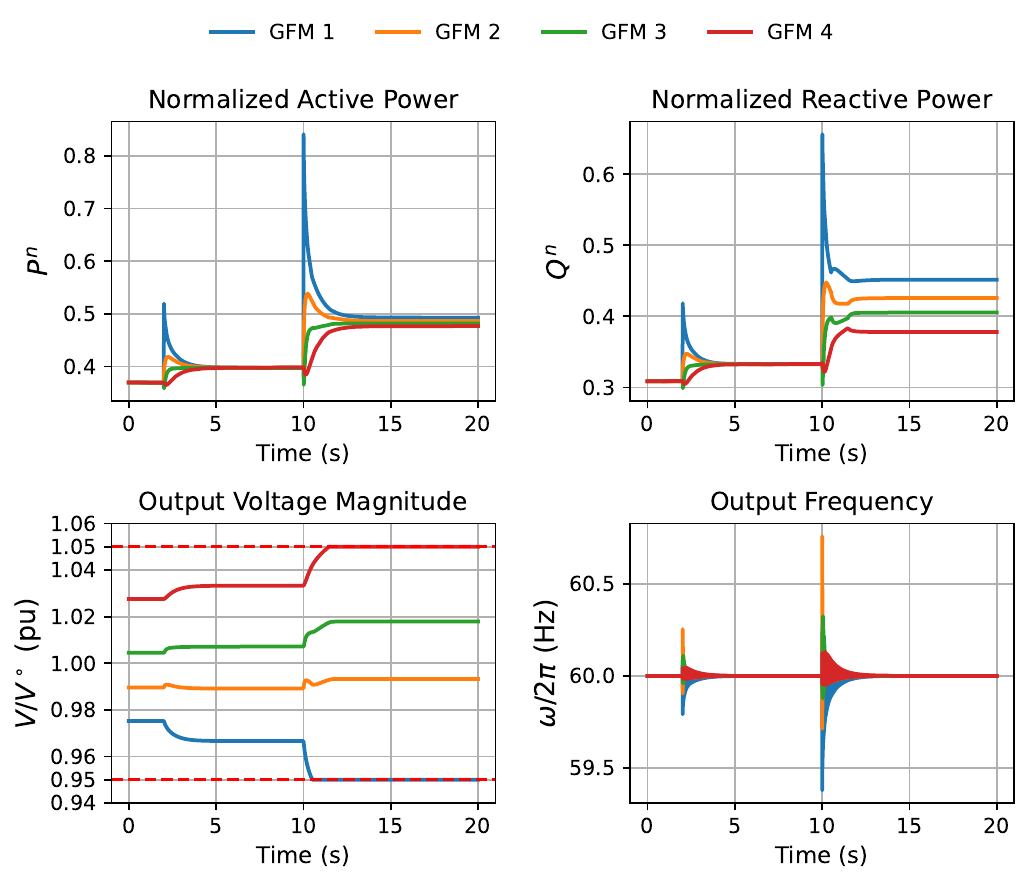}
    \caption{Normalized active and reactive power, voltage magnitude, and frequency responses under load changes at $t = 2\,\text{s}$ and $t = 10\,\text{s}$.}
    \label{fig:fat_plot_1}
\end{figure}

We consider a strong grid with line and load parameters specified in Tables~\ref{tab:line_params} and~\ref{tab:load_params}. 
A uniform gradient descent step size $\alpha_i = 0.75$ is used for all $i \in \mathcal{B}$. 
A smooth step increase in load ${\ell 1}$ is applied at $t = 2 \,\text{s}$ and $t = 10\,\text{s}$.
Fig.~\ref{fig:fat_plot_1} illustrates the system response under the proposed controller. 
Following the load change at $t = 2\,\text{s}$, the controller achieves proportional power sharing, maintains voltage magnitudes limits and restores nominal frequency within $2\,\text{s}$, thereby satisfying the three control objectives outlined in Section~\ref{sec:droop_free}.
In contrast, the disturbance at $t = 10\,\text{s}$ drives the system to its operational limits, resulting in voltage saturation at the bounds $\overline{V} = 1.05 V^\circ$ and $\underline{V} = 0.95 V^\circ$. 
Under this condition, a trade-off emerges between voltage regulation and power sharing, as the activation of voltage constraints prevents the system from achieving equal power sharing.
The proposed control ensures that normalized active and reactive powers remain as closely matched as possible while enforcing voltage limits. 

This behavior differs from the droop-free controllers proposed in~\cite{nasirian2016} and~\cite{mohiuddin2020}. 
In~\cite{nasirian2016}, only the average voltage magnitude across inverters is regulated, which does not ensure that individual voltages remain within prescribed bounds.
In~\cite{mohiuddin2020}, the voltage deviations are tightened by introducing voltage variance regulation, but done by relaxing the reactive power sharing objective through exclusion of one inverter from participation. 
The proposed method provides an intermediate approach by enforcing voltage limits while retaining participation of all inverters in the power sharing objective.

\subsection{Study 2: Transient performance in a weak grid}

To evaluate the transient performance of the proposed control relative to QSS-based droop-free methods in~\cite{nasirian2016,mohiuddinunified,mohiuddin2020}, we study the system response to a step load increase in a weak grid, with line and load parameters given in Tables~\ref{tab:line_params} and~\ref{tab:load_params}.
A step increase in load $\ell_1$ is introduced at $t = 0.2,\text{s}$, and a uniform gradient descent step size of $\alpha_i = 6.5$ is used for all $i \in \mathcal{B}$.
To ensure feasibility of the optimization problem under weak-grid conditions, where steady state voltage magnitudes may have significant variation across inverters, relaxed voltage bounds are adopted, with $\overline{V} = 1.2 V^\circ$ and $\underline{V} = 0.7 V^\circ$.

\begin{figure}
    \centering
    \includegraphics[width=\linewidth]{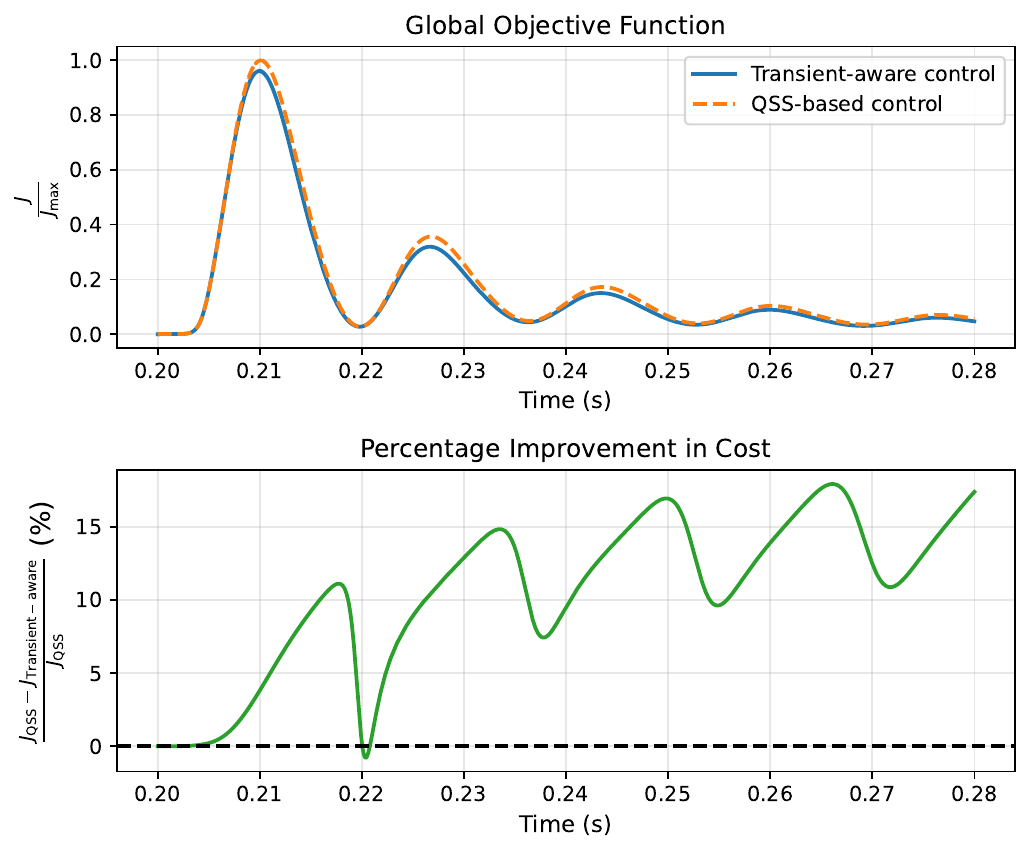}
    \caption{Global objective function for transient-aware and QSS-based control during transient response following a load change in a weak grid.}
    \label{fig:transient_cost_comparison}
\end{figure}

\begin{figure}
    \centering
    \includegraphics[width=\linewidth]{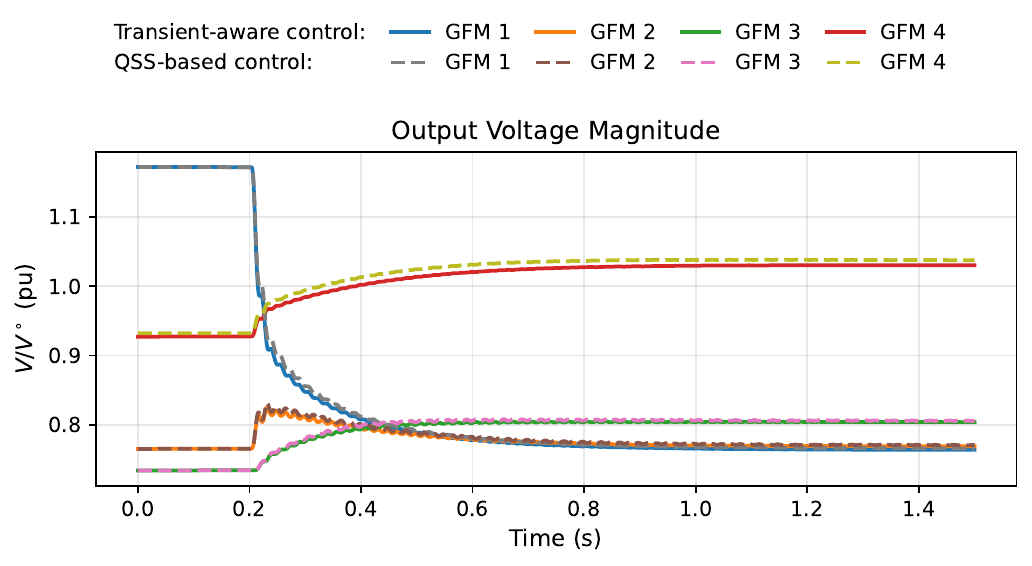}
    \caption{Voltage magnitudes under transient-aware and QSS-based control.}
    \label{fig:v_ss}
\end{figure}

Figure~\ref{fig:transient_cost_comparison} compares the transient responses of the proposed transient-aware controller and a QSS-based controller with otherwise identical objective functions and constraints. 
At steady state, both controllers achieve comparable performance, as shown in Figure~\ref{fig:v_ss}, indicating that incorporating line and load dynamics does not alter the steady state.
However, during transients, the transient-aware controller exhibits improved performance, achieving approximately $10\%$ reduction in the global objective function at peak demand compared to the QSS-based controller.
These results demonstrate that explicitly incorporating network dynamics in the control design improves responsiveness to rapid load fluctuations, particularly in weak grids where electromagnetic dynamics are more pronounced, without compromising steady state performance.

\section{Conclusion}\label{sec:conclusion}

In this paper, a transient-aware droop-free distributed primary/secondary control framework is proposed for GFM inverters in AC microgrids, with the objective of achieving proportional active and reactive power sharing, frequency regulation, and voltage regulation within prescribed bounds. 
By incorporating dynamic line and load models, the proposed controller captures electromagnetic transients neglected in QSS formulations and does not rely on decoupled active power–frequency and reactive power–voltage assumptions.
A gradient-based distributed control law is derived from an optimization framework, in which the voltage references are updated using a projection operator to enforce magnitude constraints.
The proposed formulation preserves the distributed structure of the control architecture, requiring only local measurements and limited information exchange among neighboring units.
Simulation results on a four-inverter microgrid demonstrate improved transient performance under load changes, particularly in weak grid conditions, achieving up to a 10\% improvement at peak demand compared to QSS-based droop-free methods, while preserving the same steady state operating conditions.

Future work will focus on extending the proposed framework in several directions. 
First, incorporating explicit line current constraints to ensure that thermal limits are not violated during transients. 
Second, extending the framework to more general microgrid topologies beyond inverter-only bus configurations. 
Third, improving the trade-offs between voltage regulation and power sharing by enabling more flexible utilization of inverter capacities while preserving participation among units. 
Finally, extending the transient-aware framework to systems with both GFM and GFL inverters.

\bibliographystyle{IEEEtran}
\bibliography{ref}

@ARTICLE{nasirian2016,
  author={Nasirian, Vahidreza and Shafiee, Qobad and Guerrero, Josep M. and Lewis, Frank L. and Davoudi, Ali},
  journal={IEEE Transactions on Power Electronics}, 
  title={Droop-Free Distributed Control for AC Microgrids}, 
  year={2016},
  volume={31},
  number={2},
  pages={1600-1617},
  doi={10.1109/TPEL.2015.2414457}}

@ARTICLE{mohiuddin2020,
  author={Mohiuddin, Sheik M. and Qi, Junjian},
  journal={IEEE Transactions on Smart Grid}, 
  title={Droop-Free Distributed Control for AC Microgrids With Precisely Regulated Voltage Variance and Admissible Voltage Profile Guarantees}, 
  year={2020},
  volume={11},
  number={3},
  pages={1956-1967},
  doi={10.1109/TSG.2019.2945691}}

@INPROCEEDINGS{mohiuddinunified,
  author={Mohiuddin, Sheik M. and Qi, Junjian},
  booktitle={2020 IEEE Power \& Energy Society General Meeting (PESGM)}, 
  title={A Unified Droop-Free Distributed Secondary Control for Grid-Following and Grid-Forming Inverters in AC Microgrids}, 
  year={2020},
  volume={},
  number={},
  pages={1-5},
  doi={10.1109/PESGM41954.2020.9282042}}

@INPROCEEDINGS{lasseter2002,
  author={Lasseter, R. H.},
  booktitle={2002 IEEE Power Engineering Society Winter Meeting. Conference Proceedings (Cat. No.02CH37309)}, 
  title={MicroGrids}, 
  year={2002},
  volume={1},
  number={},
  pages={305-308},
  doi={10.1109/PESW.2002.985003}}

@ARTICLE{guerrero2011,
author = {Guerrero, Josep and Vasquez, Juan C. and Alcala, Jose and Vicuna, Luis and Castilla, Miguel},
year = {2011},
month = {02},
pages = {158 - 172},
title = {Hierarchical Control of Droop-Controlled AC and DC Microgrids—A General Approach Toward Standardization},
volume = {58},
journal={IEEE Transactions on Industrial Electronics},
doi = {10.1109/TIE.2010.2066534}
}

@TECHREPORT{miller2024,
  title        = {An Overview of Grid-Forming Inverter Technologies and the Readiness of Power Systems Worldwide to Deploy the Technology},
institution  = {Electric Power Research Institute (EPRI)},
type         = {Technical Report},
  year         = {2024},
  number       = {3002031346},
  address      = {Palo Alto, CA, USA},
  author       = {N. Miller and P. Pourbeik and D. Ramasubramanian and R. Walling},
}

@ARTICLE{bahrani2024,
  author={Bahrani, Behrooz and Ravanji, Mohammad Hasan and Kroposki, Benjamin and Ramasubramanian, Deepak and Guillaud, Xavier and Prevost, Thibault and Cutululis, Nicolaos-Antonio},
  journal={IEEE Power and Energy Magazine}, 
  title={Grid-Forming Inverter-Based Resource Research Landscape: Understanding the Key Assets for Renewable-Rich Power Systems}, 
  year={2024},
  volume={22},
  number={2},
  pages={18-29},
  doi={10.1109/MPE.2023.3343338}}

@ARTICLE{bidram2012,
author = {Bidram, Ali and Davoudi, Ali},
month = {12},
year = {2012},
pages = {1963-1976},
title = {Hierarchical Structure of Microgrids Control System},
volume = {3},
journal = {IEEE Transactions on Smart Grid},
doi = {10.1109/TSG.2012.2197425}
}

@ARTICLE{pogaku2007,
  author={Pogaku, Nagaraju and Prodanovic, Milan and Green, Timothy C.},
  journal={IEEE Transactions on Power Electronics}, 
  title={Modeling, Analysis and Testing of Autonomous Operation of an Inverter-Based Microgrid}, 
  year={2007},
  volume={22},
  number={2},
  pages={613-625},
  doi={10.1109/TPEL.2006.890003}}

@ARTICLE{chandorkar,
  author={Chandorkar, M.C. and Divan, D.M. and Adapa, R.},
  journal={IEEE Transactions on Industry Applications}, 
  title={Control of parallel connected inverters in standalone AC supply systems}, 
  year={1993},
  volume={29},
  number={1},
  pages={136-143},
  doi={10.1109/28.195899}}

@ARTICLE{haddadi2014,
  author={Haddadi, Aboutaleb and Yazdani, Amirnaser and Joós, Géza and Boulet, Benoit},
  journal={IEEE Transactions on Power Delivery}, 
  title={A Gain-Scheduled Decoupling Control Strategy for Enhanced Transient Performance and Stability of an Islanded Active Distribution Network}, 
  year={2014},
  volume={29},
  number={2},
  pages={560-569},
  doi={10.1109/TPWRD.2013.2278376}}

@ARTICLE{inductiveload,
  author={Guerrero, J.M. and de Vicuna, L.G. and Matas, J. and Castilla, M. and Miret, J.},
  journal={IEEE Transactions on Power Electronics}, 
  title={A wireless controller to enhance dynamic performance of parallel inverters in distributed generation systems}, 
  year={2004},
  volume={19},
  number={5},
  pages={1205-1213},
  doi={10.1109/TPEL.2004.833451}}

@INPROCEEDINGS{shafiee2014,
  author={Shafiee, Qobad and Nasirian, Vahidreza and Guerrero, Josep M. and Lewis, Frank L. and Davoudi, Ali},
  booktitle={IECON 2014 - 40th Annual Conference of the IEEE Industrial Electronics Society}, 
  title={Team-oriented adaptive droop control for autonomous AC microgrids}, 
  year={2014},
  volume={},
  number={},
  pages={1861-1867},
  doi={10.1109/IECON.2014.7048755}
  }

@ARTICLE{adaptive1,
  author={Mahmood, Hisham and Michaelson, Dennis and Jiang, Jin},
  journal={IEEE Transactions on Smart Grid}, 
  title={Reactive Power Sharing in Islanded Microgrids Using Adaptive Voltage Droop Control}, 
  year={2015},
  volume={6},
  number={6},
  pages={3052-3060},
  doi={10.1109/TSG.2015.2399232}}

@INPROCEEDINGS{nonlinearload,
  author={Jensen, U.B. and Blaabjerg, F. and Enjeti, P.N.},
  booktitle={Conference Record of the 2000 IEEE Industry Applications Conference. Thirty-Fifth IAS Annual Meeting and World Conference on Industrial Applications of Electrical Energy (Cat. No.00CH37129)}, 
  title={Sharing of nonlinear load in parallel connected three-phase converters}, 
  year={2000},
  volume={4},
  number={},
  pages={2338-2344 vol.4},
  doi={10.1109/IAS.2000.883151}}

@ARTICLE{adaptive2,
  author={Mohamed, Yasser Abdel-Rady Ibrahim and El-Saadany, Ehab F.},
  journal={IEEE Transactions on Power Electronics}, 
  title={Adaptive Decentralized Droop Controller to Preserve Power Sharing Stability of Paralleled Inverters in Distributed Generation Microgrids}, 
  year={2008},
  volume={23},
  number={6},
  pages={2806-2816},
  doi={10.1109/TPEL.2008.2005100}}

@ARTICLE{fluxdroop,
  author={Hu, Jiefeng and Zhu, Jianguo and Dorrell, David G. and Guerrero, Josep M.},
  journal={IEEE Transactions on Power Electronics}, 
  title={Virtual Flux Droop Method—A New Control Strategy of Inverters in Microgrids}, 
  year={2014},
  volume={29},
  number={9},
  pages={4704-4711},
  doi={10.1109/TPEL.2013.2286159}}

@ARTICLE{virtualimp1,
  author={De Brabandere, Karel and Bolsens, Bruno and Van den Keybus, Jeroen and Woyte, Achim and Driesen, Johan and Belmans, Ronnie},
  journal={IEEE Transactions on Power Electronics}, 
  title={A Voltage and Frequency Droop Control Method for Parallel Inverters}, 
  year={2007},
  volume={22},
  number={4},
  pages={1107-1115},
  doi={10.1109/TPEL.2007.900456}}

@ARTICLE{virtualimp2,
  author={Guerrero, Josep M. and Vasquez, Juan C. and Matas, Jose and Castilla, Miguel and Garcia de Vicuna, Luis},
  journal={IEEE Transactions on Industrial Electronics}, 
  title={Control Strategy for Flexible Microgrid Based on Parallel Line-Interactive UPS Systems}, 
  year={2009},
  volume={56},
  number={3},
  pages={726-736},
  doi={10.1109/TIE.2008.2009274}}

\end{document}